# Resource and Application Models for Advanced Grid Schedulers

Aleksandar Lazarevic, Lionel Sacks

University College London
Email: {a.lazarevic | l.sacks @ee.ucl.ac.uk}

**ABSTRACT:** As Grid computing is becoming an inevitable future, managing, scheduling and monitoring dynamic, heterogeneous resources will present new challenges. Solutions will have to be agile and adaptive, support self-organization and autonomous management, while maintaining optimal resource utilisation. Presented in this paper are basic principles and architectural concepts for efficient resource allocation in heterogeneous Grid environment.

## 1 INTRODUCTION

Complexity of todays scientific and business problems calls for a considerably different approach to scalable computational resources. Current ground-breaking research requires unprecedented computational power and storage capacities, but current networking and administration techniques can hardly scale to such proportions. It has been recognized that integration across different platforms and administrative domains must be simplified, and resource management techniques adequately adapted to enable the system to scale better, require less human intervention, and be more resilient to hardware, software or management failures. Grid computing initiative took lead in establishing a standard for the new age of networked computing. Basic pillars of the Grid are transparent security through the use of public-key encryption (PKI), simplified administration across organizational boundaries through the use of single sign-on and policy-based management, and modular, open-standards abstraction layer that is easily implemented on heterogeneous platforms.

Globus Toolkit[1], de-facto standard reference implementation of Grid computing, implements the basic building blocks, serving more as a proof of concept then a tool for deploying a production Grid. In this paper, motivated by the need for a more robust and scalable Grid scheduler, we present a model for profiling computational resources and application requirements, and look ahead to the novel scheduling algorithms and resource matching and discovery protocols that will make use of it. We discuss ways of effectively using this model to maximise resource utilisation, provide acceptable levels of service to the user, and support a business case through Service Level Agreement (SLA) enforcement.

The proposed model is within the scope of the EPSRC funded e-Science project SO-GRM, and was developed in support, and in collaboration with other e-Science projects at UCL and in the UK. Model and scheduling logic will be based on XML and implemented using Java and Open Grid Services Architecture (OGSA)[2]. In-situ testing will be done by deploying a meta-scheduler for a production Grid cluster based on Globus Toolkit 3.

## 2 COMPUTATIONAL RESOURCE PROFILE

Previous works in this field have considered using RISC cycles, or "Deus ex machina" approaches such as Application Level Scheduling (AppLeS)[3]. These methods either require considerable amount of user interaction, application recompiling, or are inflexible for use on heterogeneous platforms such as the Grid. Our approach tries to decouple and separately model application load and node's computational output. Measurements obtained in this way are portable and can be scaled to provide short-term predictions on the availability of resources and adherence to agreed turnaround times.

However, a trade-off must be made between the complexity off the model and its accuracy. Trying to use DMTF Common Information Model [4] may prove too complex, while relying on simple scalar metrics such as MIPS or FLOPS may be over-simplistic and inaccurate.

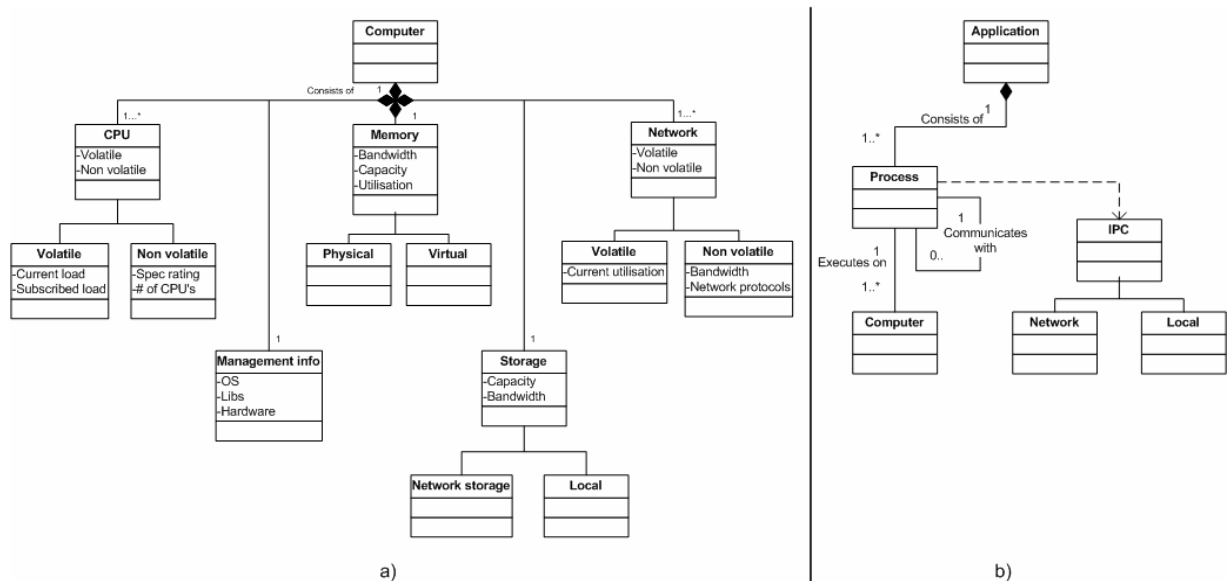

Figure 1. a) Computer Profile b) Application Profile

Figure 1(a) shows a proposed UML class diagram used to profile a computer platform on the Grid. The model is based around SPEC® cpu2000 [5] benchmark suite to assess the computational output of a node, and on simpler component-level benchmarks and information providers to capture characteristics of other performance influencing components. SPEC® suite was selected due to its credibility in the research community and widely available and comparable results. The test is distributed as ANSI C source code and readily compiles on many platforms. Test routines are thoroughly vetted and represent a cross-section of real scientific applications, with meaningful input data sets. A clear distinction is made between volatile and non-volatile resources and these two categories are separately handled at the resource matching and discovery stage. The profile exists for each machine in the Grid environment and is populated with information at runtime. At regular time intervals, volatile data is re-sampled and the profile is updated.

## 3 APPLICATION PROFILE

Figure 1(b) shows an application profile diagram. It has been developed to represent current programming practices in the e-Science community, with applications of differing architectures and varying levels of inter-process communication (IPC).

Profile creation is transparent to the user. Each application is uniquely identified by the hash-key value, and is constantly monitored during its first runtime. Profile generation and information collection are separated for portability purpose, the later using platform's native information providers (UNIX /proc system or Windows API calls for example) to obtain non-intrusive measurements.

Generally, the application execution time and resource requirements will differ on each execution, depending on the execution path the application takes, on the initial parameters, or input data sets. As part of the continuous system monitoring, these varying execution times will be recorded and used to improve the initial profile information. This statistical model will provide a measure of confidence that a job will finish within a given time/resource constraint. This feedback loop will improve the profile quality and gradually enable the reduction of overall safety margin on the cluster loading.

# 4 RESOURCE MATCHING AND DISCOVERY

Current implementations of Grid infrastructure rely on a centralised information provider (Meta Computing Directory Service – MDS) for all volatile and non-volatile aspects of the Grid state. While this might be acceptable and practical for isolated Grid "islands", the non-scalability of this approach has already been established. Our research is focused toward a distributed, self-organised system in which the information on the volatile state of the Grid is contained "in the network", rather then in any centralised database. To this effect, we will endeavour to develop a distributed meta-scheduler based on the Globus Toolkit, able to discover the current state of the resources in the Grid, and (near) optimally match application requirements to the available resources. The scheduler will be built on Open Grid Services Architecture (OGSA), and will use the information contained in the application and computer XML profiles. Resource matching will be done on non-volatile requirements first. These might include such parameters as operating system, dynamic libraries, supporting applications, minimal amounts of physical memory required, or specialised hardware. This pruned set of resources will then be passed to Self-Organized Resource Discovery protocol (SORD)[6]. Considering the application requirements as stated in the profiles, and the maximum possible computational output of each of the machines, SORD will query several close neighbours and few distant nodes to discover the optimal node for application execution.

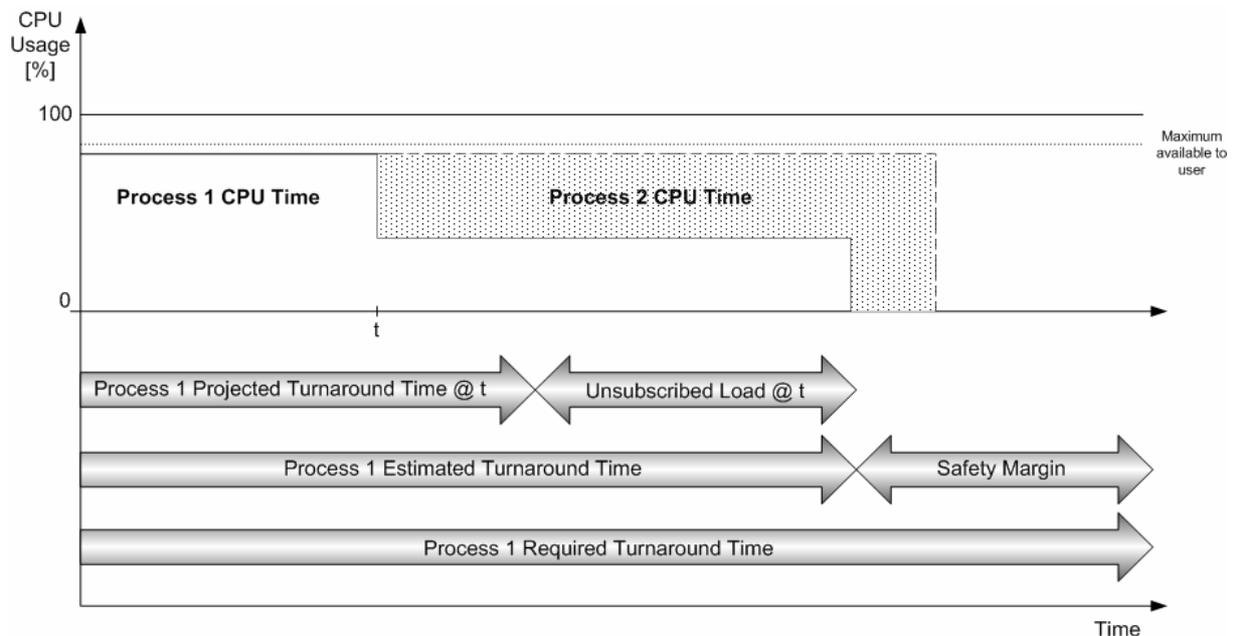

Figure 2. Effect of OS level scheduling and short-term turnaround time prediction

We introduce the "subscribed load" paradigm to circumvent the volatility of low-level scheduling in multi-tasking OS running on commodity-hardware. These systems employ soft-limiting schedulers without the ability to exclusively reserve or limit the amount of CPU time any single process is utilising. This enables statistical multiplexing of a number of concurrent processes, but introduces uncertainty in the percentage of CPU time one application will get. As shown in Figure 2. each machine maintains a log of its "subscribed load", defined as SPEC® cpu2000 marks required to execute a process, and a due time for its completion. Scheduling system will then rely on this metric rather then on spot CPU load to asses the remaining available capacity of the machine. For example, although CPU load of a machine can be very high running a single user process, the turnaround time requested might be sufficiently long to enable the process to finish with only a third of current resource utilisation. Rather than using hard-limits, operating system task scheduler is left to optimally slice CPU time, and maintain as high utilisation factor as possible. As an answer to the SORD query, a machine would answer with a specific "bid value" made up of the unsubscribed load for a given time period, and a statistically derived confidence level that running processes will actually finish within expected (profiled) time.

## 5 RESOURCE MONITORING

Monitoring the state of volatile resources in a large cluster is in itself a very resource-consuming task. Current approaches offer a centralised, database or directory-driven repository which requires regular, bandwidth consuming updates. A trade off is thus made between the volume of network traffic and the currency of information. Our approach to network monitoring is motivated by three basic requirements: reliable accounting information, low-overhead volatile resources state information for resource discovery, and low-latency data for local node integrity monitoring and process profiling.

Timely and accurate accounting information needs to be provided to the top management layers to monitor and enforce service level agreements and usage policies. This summarized information needs to be passed relatively infrequently, such as on job admission and completion, or on policy updates. To ensure reliability of data collection and its consistency, a directory service (such as MDS) or a similar structured database will be used to hold this information.

Monitoring of volatile resources in the cluster will be done using light-weight probing systems such as Ganglia and Network Weather Service (NWS). Running on each node and using IP broadcast packets to disseminate current state information, they will provide a measurement resolution of several seconds. Together with the "subscribed load" parameter, this information will affect the "bid value" presented to the SORD resource discovery protocol.

To ensure reliable node operation, and to prevent malicious or erroneous resource utilisation, an advanced integrity monitoring system has been developed. Integrity, Intelligence and Information ($I^3$)[7] system is a Java applet running on the local node and pro-actively monitoring process activity. It relies on reliable and low-latency monitoring information of key volatile system resources such as CPU and memory utilisation to detect unusual process behaviour.

By forming such three-tier information flow structure, we can ensure that critical accounting information is reliably delivered and stored, while monitoring of volatile resources remains distributed and suited for autonomous and resilient operation.

## 6 CONCLUSIONS

Effective utilisation of future global Grid environments will depend on adaptive, light-weight and autonomous resource discovery, allocation and management. Dynamic resource availability in the Grid will require scheduling algorithms to be self-organizing and self-healing, hardened against hardware and software failures. In this paper, we have presented an open-standards model for characterising computational nodes and application requirements. We have described how it can be efficiently utilised to improve scheduling in the Grid cluster and assist in monitoring resource utilisation and node integrity. Further work will focus on developing a meta-schedule for Globus Toolkit and concept testing in a production grid supporting numerous e-Science projects.

**References:**
[1] I. Foster, C. Kesselman and S. Tuecke "The Anatomy of the Grid: Enabling Scalable Virtual Organizations," *International J. Supercomputer Applications* vol. 15(3), 2001.
[2] Foster, C. Kesselman, J. Nick and S. Tuecke "The Physiology of the Grid: An Open Grid Services Architecture for Distributed Systems Integration," *Global Grid Forum* 2002.
[3] F. Berman, R. Wolski, S. Figueira, J. Schopf and G. Shao "Application-level scheduling on distributed heterogeneous networks," *Supercomputing* vol. '96, 1996.
[4] "CIM System Model White Paper," 2003, http://www.dmtf.org/standards/documents/CIM/DSP0150.pdf .
[5] J.L. Henning "SPEC® CPU2000: Measuring CPU Performance in the New Millennium," *Computer* vol. 2000, no. July, 2000.
[6] I. Liabotis, O. Prnjat, T. Olukemi, A.L. Ching, A. Lazarevic, L. Sacks, M. Fisher and P. McKee "Self-organising management of Grid environments," *International Symposium on Telecommunications* 2003.
[7] O. Prnjat, T. Olukemi, I. Liabotis and L. Sacks "Integrity and Security of the Application Level Active Networks," *IFIP Workshop on IP and ATM Traffic Management* 2001.